\documentstyle[12pt]{article}

\makeatletter
\@addtoreset{equation}{section}

\makeatother

\def\beq{\begin{equation}}
\def\eeq{\end{equation}}
\def\bea{\begin{eqnarray}}
\def\nn{\nonumber \\ }
\def\eea{\end{eqnarray}}
\def\ds{\displaystyle}
\def\sz{\scriptsize}
\def\ni{\noindent}

\def\req#1{(\ref{#1})}
\def\ie{{\it i.e.}\ }

\begin{document}

\begin{center}
{\Large\bf Improved Effective Potential in Curved Spacetime and
Quantum Matter - Higher Derivative Gravity Theory}
\\

\vspace{0.5cm}

\renewcommand{\thefootnote}{1}

{\bf Emilio Elizalde}
\footnote{E-mail: eli@zeta.ecm.ub.es. Address june-september 1994:
Department of Physics, Faculty of Sciences, Hiroshima University,
Higashi-Hiroshima 724, Japan.
E-mail: elizalde@aso.sci.hiroshima-u.ac.jp.}
\\
Center of Advanced Studies, CSIC,
Cam\'{\i} de Sta. B\`arbara, 17300 Blanes, \\
Department E.C.M. and I.F.A.E.,
Faculty of Physics, University of  Barcelona, \\
Diagonal 647, 08028 Barcelona,
Catalonia \\

\renewcommand{\thefootnote}{2}

\vskip 0.5truecm
{\bf Sergei D. Odintsov}
\footnote{On leave of absence from Tomsk Pedagogical Institute,
634041 Tomsk, Russian Federation. E-mail: odintsov@ebubecm1.bitnet }
and {\bf August Romeo},
\\
Department E.C.M., Faculty of Physics,
University of  Barcelona, \\  Diagonal 647, 08028 Barcelona,
Catalonia \\

\end{center}

\vspace{0.75cm}

\noindent{\large\bf Abstract.} We develop a general formalism to study
the renormalization group (RG) improved effective potential for
renormalizable gauge theories ---including matter-$R^2$-gravity---
in curved spacetime. The result is given up to quadratic terms in
curvature, and one-loop effective potentials may be easiliy obtained
from it. As an example, we consider scalar QED, where dimensional
transmutation in curved space and the phase structure of the potential
(in particular, curvature-induced phase trnasitions), are discussed.

For scalar QED with higher-derivative quantum gravity (QG), we examine
the influence of QG on dimensional transmutation and calculate QG
corrections to the scalar-to-vector mass ratio. The phase structure of
the RG-improved effective potential is also studied in this case, and
the values of the induced Newton and cosmological coupling constants at
the critical point are estimated. Stability of the running scalar
coupling in the Yukawa theory with conformally invariant
higher-derivative QG, and in the Standard Model with the same
addition, is numerically analyzed. We show that, in these models, QG
tends to make the scalar sector less unstable.


\begin{center}
\end{center}

\newpage

\section{Introduction}

It is common belief in modern cosmology that, during its evolution,
the Universe went through one or more inflationary stages (for an
introduction, see \cite{KT,B} and refs. therein). Some of the models
of inflationary Universe, in particular the so-called `new inflation'
one \cite{A}, are based on the scalar field effective potential
\cite{W}-\cite{J} calculated in flat space, and the corresponding phase
transitions are very important in such models. The basic observation is,
then, that curvature corrections are not too essential in the effective
potential for applications in the inflationary epoch.
However, due to the fact that we consider curved spacetimes,
for consistency one obviously has to work with curved spacetime
effective potentials.

Furthermore, it seems that inflationary universe models based on
Coleman-Weinberg type phase transition \cite{CW} are not quite
consistent (usually, the one-loop approach is used).
In these circumstances, it may happen to be very useful to
analyze the effective potential in curved spacetime (for a general
review, see \cite{BOS}) beyond one-loop approximation, in order to find
more reliable forms of it. It is well-known that effective
potentials in curved spacetimes can produce curvature-induced
phase-transitions \cite{BOS} which may become relevant in different
contexts of cosmology.

{}From another viewpoint, it would be of interest to include also in such
analysis quantum gravitational corrections. Even in the absence of a
consistent QG theory, we may work with some effective model for QG,
which can be the Einstein theory \cite{DW,HV}, which is not
renormalizable \cite{HV}.
or higher-derivative
theory \cite{BOS,S,JT} at scales below
$\mu_{\mbox{\sz Pl}} \simeq 10^{19}$ GeV. Despite its perturbative
non-unitarity, higher-derivative QG may be considered as some effective
theory, while the problem of unitarity should be addressed in a
more complete and fundamental theory. Of course, the physics which can
be adressed in that framework is between
$\mu_{\mbox{\sz GUT}} \simeq 10^{15}$ GeV and $\mu_{\mbox{\sz Pl}}$.

In the present paper we develop a general formalism to study the
RG-improved effective potential (and also the one-loop one)
for gauge theories in curved spacetime,
and also in gauge theories interacting with higher-derivative QG. The
RG-improved effective potential, which gives the leading-log behaviour
on the whole perturbation theory and hence goes beyond the one-loop
approximation, is quite well-known in flat-space theories after the
seminal paper by Coleman and Weinberg \cite{CW} (see also
\cite{EJ,FJSE}), and has numerous applications. The conception of the
RG-improved effective potential may be extended to curved spacetime
\cite{EO} and, as we shall show below, to the situation where we have an
interacting renormalizable
quantum matter-gravity theory.
We discuss a few relevant phenomena caused by the RG-improved
effective potential.

The paper is organized as follows. In the next section, our general
formalism for obtaining the RG-improved effective potential in gauge
theories on curved backgrounds ---up to quadratic terms in curvature---
is presented. One-loop effective potentials may be derived from it after
expanding for small RG-parameter $t$. This formalism is easily applied
when, in addition to matter, we have QG (any matter-gravity unified
theory should, of course, be renormalizable in our context).
In Sect. 3, the RG-improved effective potential
---in linear curvature approximation--- is explicitly given for scalar
QED. Dimensional transmutation in curved space and curvature corrections
to the scalar-to-vetor mass ratio are discussed. The phase structure of
the potential is numerically investigated for some choices of the
parameters. Sect. 4 is devoted to the study of the RG-improved (and
one-loop) effective potentials in scalar QED with higher-derivative QG.
Gravity corrections to the scalar-to-vector mass ratio are calculated
for two versions of QG (one of which is conformally-invariant
higher-derivative QG). The influence of QG on the stability of the
effective potential is numerically discussed. Spontaneous symmetry
breaking and curvature-induced phase transitions are shown to exist for
some choice of the theory parameters, and induced values of the Newton
and cosmological constants are estimated. In Sect. 5 we consider the
stability of the Yukawa theory with conformally invariant $R^2$ gravity.
Numerical analysis shows that QG corrections change the running of the
scalar coupling constant making it less unstable.
By way of some speculation,
we repeat such a discussion in the Standard Model interacting with the same
QG theory. An increase in the initial value of the QG coupling constant
can overcome the instability of the scalar coupling, which, in turn, may
change the bounds
between the Higgs and top quark masses. Finaly, we end by giving a
summary and some outlook.

\section{Renormalization-group improved effective potential in curved
spacetime}

We begin with the presentation of a general formalism for the
RG-improved effective potential. Our starting point will be some
multiplicatively renormalizable theory on a general curved background.
In principle, such a theory may include quantum gravity (QG) as well;
then, the curved background is simply the background part of the
metric (we use the background field method here).

First of all, let us consider some matter theory in curved spacetime,
where the Lagrangian corresponding to a multiplicatively renormalizable
model reads
\beq {\cal L}={\cal L}_m+{\cal L}_{\mbox{\sz ext}}, \label{L} \eeq
with
\beq
{\cal L}_{\mbox{\sz ext}}=\Lambda+ \kappa R +
a_1 R^2 + a_2 C_{\mu \nu \alpha \beta} C^{\mu \nu \alpha \beta}
+a_3 G + a_4 \Box R.
\label{Lext}
\eeq
$C_{\mu \nu \alpha \beta}$ is the Weyl tensor and $G$ is the
Gauss-Bonnet invariant. The Lagrangian for matter contains gauge fields,
some multiplets of scalars $\varphi$ and spinors $\psi$ and kinds of
interaction which are typical of any GUT. Symbolically,
\bea
{\cal L}_m&=&\ds -{1 \over 4}G_{\mu \nu} G^{\mu \nu}
+ \bar\psi (i \gamma^{\mu}{\cal D}_{\mu}-h\varphi-M ) \psi \nn
&&\ds +{1 \over 2} g^{\mu \nu} \nabla_{\mu} \varphi \nabla_{\nu} \varphi
-{1 \over 2}m^2\varphi^2 +{1 \over 2}\xi R \varphi^2
-{1 \over 4!} f \varphi^4,
\eea
where ${\cal D}_{\mu}=\nabla_{\mu}-igA_{\mu}$, $\nabla_{\mu}$ is the
covariant derivative and all indices are suppressed. Note that the
necessity of ${\cal L}_{\mbox{\sz ext}}$ is dictated by the condition
of multiplicative renormalizability in curved space \cite{BOS}. Due to
the fact that we are considering an external gravitational field,
also total derivative terms have to be included in
${\cal L}_{\mbox{\sz ext}}$.

We will be interested in the calculation of the effective potential
\cite{W}-\cite{J} for the scalar field, \ie the effective action on a
constant background $\varphi, R$. Since the theory is multiplicatively
renormalizable, its effective potential satisfies the standard RG
equation:
\beq
\left(
 \mu {\partial\over \partial\mu}
+\beta_i {\partial\over \partial\lambda_i}
-\gamma\varphi {\partial\over \partial\varphi}
\right)
V( \mu, \lambda_i, \varphi ) =0,
\label{RGeqV}
\eeq
where
$\lambda_i=
( \xi, f, h^2, g^2, m^2, \mu, \Lambda, \kappa, a_1, a_2, a_3)$,
$\beta_i$ are the corresponding beta functions, and $\varphi$ is the
background scalar. The $\Box R$ term from \req{Lext} has disappeared
owing to our choice of background space of the form $R=$ const. For
the gauge fields, a Landau-type gauge is supposed to be chosen.
As a result, in the one loop approach, which we apply for handling
\req{RGeqV}, the
beta function for the gauge parameter is zero and its associated term in
\req{RGeqV} may be dropped out. The RG equation for $V$
has been discussed in \cite{CW,EJ,FJSE} (and refs. therein) for flat space
and in \cite{BOS,EO} for the case of curved space.

Solving eq. \req{RGeqV} by the method of the characteristics gives
\beq
V( \mu, \lambda_i, \varphi )=
V( \mu e^t, \lambda_i(t), \varphi(t) ),
\label{Vft}
\eeq
where
\beq
\begin{array}{lllllll}
\ds{d\lambda_i (t) \over dt}&=&\beta_i(\lambda_i(t)),&\hspace{1cm}&
\lambda_i(0)&=&\lambda_i, \\
\ds{d\varphi (t) \over dt}&=&-\gamma(t)\varphi(t),&\hspace{1cm}&
\varphi(0)&=&\varphi.
\end{array}
\eeq
Physically, \req{Vft} means that the effective potential can be
(locally) found provided that its functional at some certain $t$ is
known.

Using the classical Lagrangian \req{L} for constant background as the
initial value of $V$ at $t=0$, and working only in one-loop
aproximation, we can find \cite{EO}
\bea
V_{\mbox{\sz RG}}&=&\ds{1 \over 4!}f(t) \varphi^4(t)
-{1 \over 2} \xi(t) R \varphi^2(t)+{1 \over 2} m^2(t) \varphi^2(t) \nn
&&\ds +\Lambda(t) + \kappa(t) R + a_1(t)R^2
+a_2(t)C_{\mu \nu \alpha \beta} C^{\mu \nu \alpha \beta} +a_3(t) G,
\label{VRGwithas}
\eea
where
\beq
\begin{array}{lllllllll}
\dot f(t)&= \beta_f(t),&f(0)&=f,&\hspace{1cm}&
\dot\kappa(t)&=\beta_{\kappa}(t),&\kappa(0)&=\kappa, \\
\dot\xi(t)&=\beta_{\xi}(t),&\xi(0)&=\xi,&\hspace{1cm}&
\dot a_i(t)&=\beta_{a_i}(t),&a_i(0)&=a_i, \ i=1,2,3, \\
\dot m^2(t)&=\beta_{m^2}(t),&m^2(0)&=m^2,&\hspace{1cm}&
\dot\varphi(t)&=-\gamma(t)\varphi(t),&\varphi(0)&=\varphi, \\
\dot\Lambda(t)&=\beta_{\Lambda}(t),&\Lambda(0)&=\Lambda,&&&&&
\end{array}
\label{viibetas}
\eeq
The one-loop beta functions which appear on the r.h.s. of \req{viibetas}
for any particular model are, as a rule, known, or may be obtained
without serious problems. It should be observed that the RG-improved
effective potential \req{VRGwithas} was written in the approximation
up to curvature invariants of second order.

This RG-improved efective potential (EP) is given in
leading-log approximation (summing all leading logs in perturbation
theory) \cite{CW}, which, in this sense, is much richer than the
standard one-loop version. One loop EPs can be obtained from RG-improved
EPs in some limit (small $t$, weak couplings). However, notice that,
contrary to what happens in the non-improved case, eq. \req{VRGwithas}
is actually valid at all $t$'s for which $V_{\mbox{\sz RG}}$
does not diverge (this is an improvement).

Now, we ask ourselves this natural question: what is the choice
of the RG parameter $t$ which leads to the summation of all logarithms
to all orders? In fact, for massive theories it is not easy to answer
\cite{FJSE,EO}. One has to introduce a few massive scales (which make
the discussion technically complicated), use the decoupling theorem
\cite{AC} and the effective field theory \cite{Wetal} (and refs.
therein) to construct the RG-improved EPs at all these scales. In the
present paper we shall consider, for simplicity, either massless
theories, where the choice is actually unique
$\ds t={1 \over 2}\log{ \varphi^2 \over \mu^2 }$, or massive theories
limited to the case of very high $\varphi$, such that
$\varphi^2 \gg m_{\mbox{\sz eff}}^2$, being $m_{\mbox{\sz eff}}^2$ the
largest effective mass of the theory. Then, we may drop all massive
terms in \req{VRGwithas} and make our choice of $t$, again, as above.

In massless theories the RG-improved potential \req{VRGwithas}
may be expanded for
small $t$ and weak coupling in a general form, thus obtaining a very
general expression
for the one-loop effective potential \cite{BO} (see also \cite{BOS}):
\beq
\begin{array}{lll}
V^{(1)}&=&\displaystyle{1 \over 4!} f\varphi^4
+{1 \over 48} (\beta_f -4f \gamma ) \varphi^4
\left( \log {\varphi^2 \over \mu^2 } - {25 \over 6} \right) \\
&&\displaystyle-{1 \over 2}\xi R \varphi^2
-{1 \over 4} (\beta_{\xi} -2\xi \gamma ) R \varphi^2
\left( \log {\varphi^2 \over \mu^2 } -3 \right),
\end{array}
\label{V1}
\eeq
where $\beta_f, \beta_{\xi}, \gamma$ are the one-loop beta functions,
and $\mu^2$ is a mass parameter in the range
$\mu^2 { < \atop \sim } \mu_{\mbox{\sz GUT}}^2 \simeq 10^{15}$ GeV.
Note that Coleman-Weinberg type normalization conditions have been used
to derive eq. \req{V1} (see \cite{BO} for more details). This is a very
useful expression which will be employed for explicit analysis in
a few different theories below.

Next, we would like to point out that the above developed formalism
can be easily applied to multiplicatively renormalizable QG
theories with matter. As is widely well-known, Einstein QG
is a non-renormalizable theory \cite{HV}. That is why we have chosen
to work with higher-derivative QG, which is multiplicatively
renormalizable \cite{S,BOS} and asymptotically free \cite{JT,BKSVW,BOS}.
It should be noted that it may be asymptotically free for all couplings
when such a theory interacts with some GUT model \cite{BKSVW,BOS}.
Of course, $R^2$-gravity with matter cannot be considered as
a reasonable candidate for a consistent QG theory, due to the
open question about its perturbative non-unitarity, which is typical
of any higher-derivative interaction. However, one can regard it
as an effective model for some yet unknown, consistent, QG theory
at scales below $\mu_{\mbox{\sz Pl}} \simeq 10^{19}$ GeV.
That
will be our viewpoint througout this paper.

Working, for simplicity, with massless theories, our initial Lagrangian
---using standard notations--- is
\bea
{\cal L}&=&{\cal L}_m+{\cal L}_{\mbox{\sz QG}}, \nn
{\cal L}_{\mbox{\sz QG}}&=&\ds{1 \over \lambda}W
-{\omega \over 3 \lambda} R^2,
\label{LQG}
\eea
where $W=C_{\mu \nu \alpha \beta} C^{\mu \nu \alpha \beta}$. Even in
this case, starting from higher-derivative QG without Einstein and
cosmological terms, we may recover them as a result of a
curvature-induced phase transitions at lower energies, as we will see
below. When dealing with such a theory, we get again an RG-improved EP
of the form \req{VRGwithas}, and a one-loop efective potential like
\req{V1}.
The only difference, in comparison with the no-QG case, is that all beta
functions have now changed due to explicit QG corrections. Hence,
our formalism is general enough for being applied to a QG theory also.

Is should also be observed that, as the metric is now quantized, one has
to introduce the term
$\ds -\gamma_g g_{\mu \nu} {\delta \over \delta g_{\mu \nu}}$ in
eq. \req{RGeqV} \cite{BOS}. However, in one-loop approximation and
using the background field method, $\gamma_g=0$ and therefore this term
vanishes. Having at hand the general formalism developed in this
section, we may now set out to study explicit examples.

\section{Renormalization-group improved effective potential in gauge
theories in curved spacetime}

Let us begin with the simplest model for an Abelian gauge theory:
massless electrodynamics in curved spacetime. The classical Lagrangian
for this theory is
\bea
{\cal L}_m&=&\ds{1 \over 2}\left(
\partial_{\mu}\varphi_1 -eA_{\mu}\varphi_2
\right)^2
+{1 \over 2}\left(
\partial_{\mu}\varphi_2 -eA_{\mu}\varphi_1
\right)^2 \nn
&&\ds+{1 \over 2} \xi R \varphi^2
-{1 \over 4!} f \varphi^4 -{1 \over 4}F_{\mu \nu} F^{\mu \nu},
\label{Lm}
\eea
where $\varphi^2=\varphi_1^2+\varphi_2^2$. Using Landau gauge for
calculating the one-loop beta functions, these turn out to be
\cite{CW,EO}:
\beq
\begin{array}{c}
\ds \beta_f={ 1 \over (4 \pi)^2 }
\left( {10 \over 3} f^2 -12 e^2 f +36 e^4 \right), \\
\ds \beta_{e^2}={ 2 e^4 \over 3 (4 \pi)^2 }, \hspace{0.5cm}
\gamma=-{3 e^2 \over (4\pi )^2}, \hspace{0.5cm}
\beta_{\xi}= { \left( \xi -{1 \over 6} \right) \over (4\pi )^2 }
\left( {4 \over 3} f-6e^2 \right) .
\end{array}
\label{betafsLandau}
\eeq
 From here on, limiting ourselves to linear curvature approximation, the
$\beta_{a_i}$'s will be no longer necessary in our discussion.

The solutions of the RG equations for the coupling constants are
\beq
\begin{array}{c}
\ds e^2(t)=e^2\left( 1- { 2 e^2 t \over 3 (4 \pi)^2 } \right)^{-1},
\hspace{0.5cm}
\varphi^2(t)=\varphi^2(t)
\left( 1- { 2 e^2 \over 3 (4 \pi)^2 } \right)^{-9},
\\
\vspace*{0.2cm}
\ds f(t)={1 \over 10}e^2(t)\left[
\sqrt{719} \tan \left( {1 \over 2} \sqrt{719} \log e^2(t) + C \right)
+19 \right] ,
\\
\vspace*{0.2cm}
\ds C= \arctan\left[
{1 \over \sqrt{719}}\left( {10 f \over e^2}- 19 \right)
\right]
-{1 \over 2} \sqrt{719} \log e^2,
\\
\ds \xi(t)={1 \over 6} +\left( \xi - {1 \over 6} \right)
\left[ {e^2(t) \over e^2} \right]^{-26/5}
{ \cos^{2/5} \left( {1 \over 2} \sqrt{719} \log e^2 + C \right) \over
\left( {1 \over 2} \sqrt{719} \log e^2(t) + C \right) } .
\end{array}
\eeq
Using these effective coupling constants, the RG improved potential
is found to be \cite{EO}
\beq
V={1 \over 4! } f(t) \varphi^4(t)
-{1 \over 2} \xi(t) R \varphi^2(t) .
\label{VRGiSQEDm0}
\eeq
This RG improved EP in scalar electrodynamics has already been discussed
in a more general case, namely, with a massive scalar field \cite{EO}.

Before starting to work with \req{VRGiSQEDm0}, we write down the one
loop EP \req{V1} for this specific theory:
\beq
\begin{array}{ll}
V^{(1)}=&\ds{1 \over 4!} f\varphi^4
+{1 \over 48 (4\pi)^2}
\left[ {10 \over 3} f^2 + 36e^4 \right] \varphi^4
\left( \log {\varphi^2 \over \mu^2 } - {25 \over 6} \right) \\
&\ds -{1 \over 2} \xi R\varphi^2
+{1 \over 12 (4\pi)^2}
\left[ \left( \xi-{1 \over 6} \right) \left( {4 \over 3}f -6e^2 \right)
+ 6\xi e^2 \right]
R \varphi^2 \left( \log {\varphi^2 \over \mu^2 } - 3 \right).
\label{V1SQEDnoQG}
\end{array}
\eeq
This one-loop EP in linear approximation was also obtained in refs.
\cite{Sh}, where a background De Sitter space was considered for
explicit calculations.

A few remarks concerning dimensional transmutation are in order.
Restricting $V^{(1)}$ first to flat space ($R=0$), choosing next
$\mu=\varphi_m$ where $\varphi=\varphi_m$ is the vacuum state, and
repeating
the analysis by Coleman and Weinberg \cite{CW}, one can easily get
\beq
\begin{array}{lll}
V^{(1)}&=&\ds{3 e^4 \over 64 \pi^2} \varphi^4
\left( \log {\varphi^2 \over \varphi_m^2 } - {1 \over 2} \right), \\
\ds{f \over 24}&=&\ds{11 \over 144 (4\pi)^2} (36 e^4) .
\end{array}
\label{V1andcond}
\eeq
When carrying these considerations over to curved spacetime, the
transmutation mechanism acts in a different way \cite{EO}. Indeed,
supposing again $f \sim e^4$ we arrive at
\beq
\begin{array}{ll}
V^{(1)}=&\ds{1 \over 4!} f\varphi^4
+{3 e^4 \over 64 \pi^2} \varphi^4
\left( \log {\varphi^2 \over \mu^2 } - {25 \over 6} \right) \\
&\ds -{1 \over 2} \xi R\varphi^2
-{1 \over 4 (4\pi)^2} e^2
R \varphi^2 \left( \log {\varphi^2 \over \mu^2 } - 3 \right).
\end{array}
\eeq
Choosing $\mu=\varphi_m$, where $\varphi=\varphi_m$ is now the vacuum
state in curved spacetime (supposing that it exists),
we do not find such a
precise connection between $f$ and $e^4$ as in flat space \cite{CW}
but, instead, we obtain
\beq
{ V^{(1) \prime } \over 4 \varphi_m } =
\left[ {f \over 4!} - {11 e^4 \over 64 \pi^2} \right] \varphi_m^2
+R \left[ -{1 \over 4}\xi + {e^2 \over 4(4 \pi )^2} \right]=0 .
\eeq
Working in linear curvature approximation (assuming that curvature
corrections are not large) one can impose the second condition in
\req{V1andcond} by hand. Then, and only then, is dimensional
transmutation switched on in curved space, yielding a condition to
fix $\xi$ in terms of $e^2$:
\beq \xi= {e^2 \over (4 \pi )^2} . \label{xie2} \eeq
In this case, the universal normalization independent expression
for the one-loop effective potential turns into
\beq
V^{(1)}={3 e^4 \over 64 \pi^2} \varphi^4
\left( \log {\varphi^2 \over \mu^2 } - {1 \over 2} \right)
-{e^2 \over 64 \pi^2}
R \varphi^2 \left( \log {\varphi^2 \over \mu^2 } - 1 \right) .
\label{V1withm12m1}
\eeq
Obviously, the above arguments do not hold for models in the presence of
strong curvature, where other choices leaving $\xi$ arbitrary seem to be
more natural.

By use of \req{V1withm12m1}, one can find the constant curvature
corrections to the scalar-to-vector mass ratio in the following form
(see \cite{CW}):
\beq
{m^2(S) \over m^2(V)} =
{ V^{(1) \prime \prime }(\varphi_m) \over e^2 \varphi_m^2} =
{3 e^2 \over 8\pi^2} - {R \over 16 \pi^2 \varphi_m^2} .
\label{massratioStoV}
\eeq
That gives the non-zero curved space generalization of
the corresponding flat space Coleman-Weinberg result \cite{CW}.

Our purpose now will be to discuss the phase structure, \ie symmetry
breaking and curvature-induced phase transitions, in scalar QED,
using the RG-improved effective potential. We
shall be interested in first order phase transitions, \ie those in
which the order parameter jumps sharply at some critical value of the
curvature $R_c$. The conditions on such phase transitions are \cite{Sh}
\beq
V( \varphi_c, R_c )=0, \hspace{0.5cm}
\left.{\partial V \over \partial\varphi}\right\vert_{ \varphi_c, R_c }=0,
\hspace{0.5cm}
\left.{\partial^2 V \over \partial\varphi^2}
\right\vert_{ \varphi_c, R_c } >0 .
\eeq
The behaviour of this RG-improved potential in one of the most
interesting cases ---associated to $e^2=0.1$, $\xi=0$---
is shown in Fig. 1.  If $R$ is smaller than a certain value $R_1$
around $1.9\cdot 10^{-6}$, a local minimum for
some $\varphi >0$ exists. However, that state is just metastable until
the value of $R$ is lowered down to $R_c \simeq 1.44 \cdot 10^{-6}$.
Below this
figure, the new minimum is the global one, \ie $\varphi=0$ has become
metastable while the $\varphi\neq0$ associated to the global minimum is
now the physical vacuum. Therefore, a symmetry-breaking phase
transition, induced by curvature itself, takes place at $R=R_c$, even
in a situation with $\xi(0)=0$. The fact that scalar QED
in curved spacetime undergoes a curvature-induced phase-transition
was already noted, some time ago \cite{Sh}, using the one-loop effective
potential.

By our explicit numerical results, we have shown that the
RG-improved effective potential behaves qualitatively like
its one-loop counterpart. Using the one-loop potential \req{V1SQEDnoQG}
(keeping, for simplicity, only logarithmic terms in the one loop
correction) we have found curves which differ just very slightly from
the ones plotted in Fig. 1.
Of course, it would not be difficult to repeat the same
analysis for different choices of the initial values of the theory
parameters.

\section{The effective potential in scalar QED with
higher-derivartive gravity}

We will now look at the effective potential in massless scalar QED
interacting with QG in the form \req{LQG} and \req{Lm}. QG corrections
to QED beta functions, which may be taken from ref. \cite{BKSVW} (see
also \cite{BO}), have a universal form for any matter theory.
Unfortunately, the complexity of the RG equations for the total system
of beta functions prevents us from finding the RG-improved EP
\req{VRGwithas} explicitly. It can be obtained only in an implicit form
(applying linear curvature approximation):
\beq
\ds V={1 \over 4! } f(t) \varphi^4(t)
-{1 \over 2} \xi(t) R \varphi^2(t),
\eeq
where
\beq
\begin{array}{c}
\ds \lambda(t)={ \lambda \over
\ds 1+ { \alpha^2\lambda t \over (4 \pi)^2 } }, \hspace{1cm}
\alpha^2= {203 \over 15}, \hspace{1cm}
e^2(t)={ e^2 \over
\ds 1- { 2 e^2 t \over 3 (4 \pi)^2 } },
\\
\ds{d\omega \over dt}=\beta_{\omega}=
-{1 \over (4 \pi)^2}\lambda\left[
{10 \over 3}\omega^2 +\left( 5+\alpha^2 \right)\omega +{5 \over 12}
+3\left( \xi-{1 \over 6} \right)^2 \right], \\
\ds {d\xi\over dt}= \beta_{\xi}+\Delta\beta_{\xi}, \hspace{0.5cm}
{df \over dt}=\beta_f +\Delta\beta_f, \hspace{0.5cm}
-{1 \over \varphi}{d\varphi\over dt}=
\gamma + \Delta\gamma .
\end{array}
\label{betafsSQEDQG}
\eeq
In expressions \req{betafsSQEDQG}, $\beta_{\xi}, \beta_f, \gamma$ are
the ones given in \req{betafsLandau}, and the universal QG corrections
\cite{BKSVW,BOS} are
\begin{equation}
\begin{array}{lll}
\ds\Delta\beta_{\xi}&=&\ds{1 \over (4 \pi)^2}\lambda\xi \left[
-{3 \over 2}\xi^2 +4\xi +3 +{10 \over 3}\omega
+{1 \over \omega}\left( -{9 \over 4}\xi^2 +5 \xi+ 1 \right)
\right], \\
\ds\Delta\beta_f&=&\ds{1 \over (4 \pi)^2}
\left[
\lambda^2\xi^2\left( 15+ {3 \over 4\omega^2}
-{9 \xi \over \omega^2}+{27 \xi^2 \over \omega^2} \right) \right. \\
&&\hspace{3em}\displaystyle\left. -\lambda f\left(
5+3\xi^2+{33 \over 2\omega}\xi^2
-{6 \over \omega}\xi +{1 \over 2 \omega} \right) \right], \\
\ds\Delta\gamma&=&\ds{1 \over (4 \pi)^2}
{\lambda \over 4}\left(
{13 \over 3} -8\xi -3\xi^2 -{1 \over 6 \omega} -{2 \xi \over \omega}
+{3 \xi^2 \over 2 \omega}
\right),
\end{array}
\label{Deltabetafs}
\eeq
As one can see from \req{betafsSQEDQG}, there are no QG corrections to
$e^2(t)$ (that is prohibited by local gauge invariance). The calculation
of $\Delta\gamma$ took place in harmonic gauge \cite{BKSVW}. For
simplicity, we have not explicitly written in
\req{betafsSQEDQG}, \req{Deltabetafs} the $t$-dependence. Moreover,
for these equations, the standard initial conditions
$\omega(0)=\omega$, $f(0)=f$, $\xi(0)=\xi$, $\varphi(0)=\varphi$ were
assumed (here, the $\omega$, $f$, $\xi$, $\varphi$ on the r.h.s. denote
{\it truly} $t$-independent quantities).

For the conformal version of higher-derivative gravity with scalar QED,
one has
\beq
\ds V={1 \over 4! } f(t) \varphi^4(t)
-{1 \over 12} R \varphi^2(t),
\label{VRGconf}
\eeq
where
\beq
\begin{array}{c}
\ds \lambda(t)={ \lambda \over
\ds 1+ { \alpha_1^2\lambda t \over (4 \pi)^2 } }, \hspace{1cm}
\alpha_1^2= {27 \over 2}, \hspace{1cm}
e^2(t)={ e^2 \over
\ds 1- { 2 e^2 t \over 3 (4 \pi)^2 } },
\\
\begin{array}{rcl}
\ds{df \over dt}&=&\ds
{1 \over (4 \pi)^2}\left(
{10 \over 3}f^2 -12 e^2 f + 36 e^4
-{41 \over 8}\lambda f+ {5 \over 12}\lambda^2 \right), \\
\ds -{1 \over \varphi}{d\varphi\over dt}&=&\ds
{1 \over (4 \pi)^2} \left( -3e^2 +{27 \over 32}\lambda \right) .
\end{array}
\end{array}
\eeq
and the QG corrections in this conformal model have been taken
from \cite{BKSVW}. Note that, in order to make such a theory
multiplicatively renormalizable, one has to use the so-called
special conformal regularization (see \cite{BOS,BKSVW}, third ref.
of \cite{JT} and also refs. therein).

Using the above beta functions, it is easy to find
the one loop effective potential \req{V1} for higher-derivative QG with
scalar QED. In the general version \req{LQG}, taking into account
\req{betafsSQEDQG}, \req{Deltabetafs} one can obtain
(see also \cite{OER})
\beq
\begin{array}{ll}
V^{(1)}=&\ds{1 \over 4!} f\varphi^4  \\
&\ds+{1\over 48(4 \pi)^2}\left[
{10 \over 3}f^2 +36 e^4
+\lambda^2\xi^2\left(
15+ {3 \over 4\omega^2} -{9 \xi \over \omega^2}+{27 \xi^2 \over \omega^2}
\right) \right. \\
&\ds\hspace{5em}\left. -\lambda f\left(
{28\over 3}+18{\xi^2 \over \omega}-{8 \xi \over \omega} -8\xi
+{1 \over 3 \omega} \right) \right]
\varphi^4
\left( \log {\varphi^2 \over \mu^2 } - {25 \over 6} \right) \\
&\ds -{1 \over 2}\xi R \varphi^2 \\
&\ds -{1\over 4(4 \pi)^2}\left\{ \left( \xi-{1 \over 6} \right)
\left( {4 \over 3}f- 6 e^2 \right) + 6 \xi e^2 \right. \\
&\ds\hspace{5em}\left.+\lambda\xi\left[
8\xi+{5 \over 6}+{10\over 3}\omega
+{1 \over \omega}\left( -3\xi^2 +6\xi +{13 \over 12} \right) \right]
\right\}
R \varphi^2
\left( \log {\varphi^2 \over \mu^2 } - 3 \right) .
\end{array}
\label{V1SQEDQGgv}
\eeq
In this case, we suppose that
$\mu_{\mbox{\sz GUT}}^2 < \mu^2 < \mu_{\mbox{\sz Pl}}^2$.

As for the conformal version, the one-loop EP turns out to be as
follows:
\beq
\begin{array}{ll}
V^{(1)}=&\ds{1 \over 4!} f\varphi^4
+{1 \over 48 (4\pi)^2}
\left[ {10 \over 3} f^2 + 36e^4+ {5 \over 12}\lambda^2
-{17 \over 2} \lambda f \right] \varphi^4
\left( \log {\varphi^2 \over \mu^2 } - {25 \over 6} \right) \\
&\ds -{1 \over 12}R\varphi^2
+{1 \over 12 (4\pi)^2}
\left[ -3e^2+{27 \over 32}\lambda \right]  R \varphi^2
\left( \log {\varphi^2 \over \mu^2 } - 3 \right).
\end{array}
\label{V1SQEDQGconf}
\eeq

First, we discuss the dimensional transmutation in presence of QG (see
also \cite{OER}) on flat background.
Taking eq. \req{V1SQEDQGgv}, supposing
either $\lambda^2$ and $e^4$ of the same order or $\lambda^2$ dominant,
choosing $\mu^2=\varphi_m^2$ (where $\varphi_m$ is the minimum), and
applying
$\ds \left. {\partial V^{(1)} \over \partial\varphi }
\right\vert_{\varphi=\varphi_m}=0$
as in Sect. 3, one finds that, due to dimensional transmutation
$f \sim \lambda^2 + e^4$, and therefore
\beq
\begin{array}{c}
\ds V^{(1)}={1 \over 48 (4\pi)^2}
\left[
36 e^4
+\lambda^2 \xi^2 \left(
15+ {3 \over 4\omega^2} -{9 \xi \over \omega^2}+{27 \xi^2 \over \omega^2}
\right)
\right]
\varphi^4
\left( \log {\varphi^2 \over \varphi_m^2 } - {1 \over 2} \right) . \\
\ds 2 (4 \pi)^2 f = {11 \over 3} \left[
\lambda^2\xi^2 \left(
15 + {3 \over 4\omega^2 }
-{9 \xi \over \omega^2}+{27 \xi^2 \over \omega^2}
\right)
+36 e^4 \right] .
\end{array}
\label{V1SQEDQGtransmut}
\eeq
If $e^4$ is leading as compared with the QG term, we are in the case
already studied by Coleman and Weinberg \cite{CW}: QG corrections are
negligible. However, in the opposite situation, where
$\lambda^2 \gg e^4$, dimensional transmutation is a purely quantum
gravitational effect.

 From eq. \req{V1SQEDQGtransmut} one can obtain the scalar-to-vector mass
ratio in the presence of QG
\beq
{ m^2(S) \over m^2(V) }=
{1 \over 6 (4\pi)^2}
\left[
36e^2
+{\lambda^2 \xi^2 \over e^2} \left(
15+ {3 \over 4\omega^2} -{9 \xi \over \omega^2}+{27 \xi^2 \over \omega^2}
\right)
\right] .
\label{massrtg}
\eeq
This is to be compared with the original Coleman-Weinberg \cite{CW} result
($\lambda^2=0$ case); see \req{massratioStoV}.
As one can notice, the QG corrections
in \req{massrtg} may become the dominant ones.

We also find the analogous ratio for the conformal version, which
reads
\beq
{ m^2(S) \over m^2(V) }=
{1 \over 6 (4\pi)^2}
\left[ 36e^2 +{5 \over 12} {\lambda^2  \over e^2} \right] .
\label{massrtc}
\eeq
Again, QG corrections may turn out to be the leading ones.

It is interesting to realize that, in principle, curvature corrections
to \req{massrtg} or \req{massrtc} may be calculated as we did in Sect.
3.
Next, our aim is to define $\xi$ in terms of $e^2$, $\lambda$ in the
general version.
To this end, we
take \req{V1SQEDQGgv} assuming $f \sim \lambda^2+e^4$, and get:
\beq
\begin{array}{ll}
V^{(1)}=&\ds{1 \over 4!} f\varphi^4  \\
&\ds+{1\over 48(4 \pi)^2}\left[
36 e^4
+\lambda^2\xi^2\left(
15+ {3 \over 4\omega^2} -{9 \xi \over \omega^2}+{27 \xi^2 \over \omega^2}
\right) \right]
\varphi^4
\left( \log {\varphi^2 \over \mu^2 } - {25 \over 6} \right) \\
&\ds -{1 \over 2}\xi R \varphi^2 \\
&\ds -{1\over 4(4 \pi)^2}
\left\{
e^2
+\lambda\xi\left[
{5 \over 6}+8\xi+{10\over 3}\omega
+{1 \over \omega}\left( -3\xi^2 +6\xi +{13 \over 12} \right) \right]
\right\}
R \varphi^2
\left( \log {\varphi^2 \over \mu^2 } - 3 \right) .
\end{array}
\label{V1SQEDQGgvwoxi}
\eeq
Choosing $\mu=\varphi_m$ and requiring
$\ds \left. {\partial V^{(1)} \over \partial\varphi }
\right\vert_{\varphi=\varphi_m}=0$ on \req{V1SQEDQGgvwoxi}, we do not
obtain the same connection among $f$, $\lambda^2$ and $e^4$ as on flat
background (the situation is the same as in Sect. 2). If we impose such
condition, following
from flat space considerations, we get the expression \req{xie2}, as the
connection between $\xi$ and $e^2$, even in the presence of QG.
This fact is caused by the particular form of the QG corrections to
the $R \varphi^2$ term in \req{V1SQEDQGgvwoxi} (it is always
proportional to $\xi$). It is also very interesting to note that,
if we started from the theory without electrodynamic sector, \ie
$e^2=0$, we would get, from the condition
$\ds \left. {\partial V^{(1)}(\mu = \varphi_m)
\over \partial\varphi} \right\vert_{\varphi_m} = 0$, the equation
\[
\varphi_m^2 \left\{ {f \over 6}
-{11 \over 3} \cdot {1 \over 12 (4 \pi)^2 }
\lambda^2\xi^2\left(
15+ {3 \over 4\omega^2} -{9 \xi \over \omega^2}+{27 \xi^2 \over \omega^2}
\right) \right\}
\]
\beq
-R \left\{ \xi - {1 \over (4 \pi)^2 }
\lambda\xi\left[
{5 \over 6}+8\xi+{10\over 3}\omega
+{1 \over \omega}\left( -3\xi^2 +6\xi +{13 \over 12} \right) \right]
\right\} =0 .
\eeq
Generally speaking, this is an equation to determine the minimum
(assuming that it exists) in terms of the curvtaure and the theory
parameters. However, if ---as before--- we put by hand the flat space
condition for $f$, we are led to
\beq
1 - {\lambda \over (4 \pi)^2 }
\left[
{5 \over 6}+8\xi+{10\over 3}\omega
+{1 \over \omega}\left( -3\xi^2 +6\xi +{13 \over 12} \right) \right]
=0.
\eeq
This condition is inconsistent, as it causes one of the QG coupling
constants to be larger than unity, which contradicts the implicit
assumptions in our perturbative treatment of the theory. Hence, without
the electromagnetic coupling constant $e$, dimensional transmutation
does not work order by order in curvature.

Some properties of the model under consideration are illustrated by
means of Figs. 2-5.
The scalar coupling $f(t)$ is represented in Fig. 2, for the conformal
version, and in Fig. 3, for a case of the general model where
this function is seen to go negative in a certain region, thus
turning the effective potential unstable in that range.
While
conformal invariance seems to prevent instabilities of this type,
in a general situation the positiveness of this coupling cannot be taken
for granted.

As for the chances of symmetry breaking,
the appearance of a global minimum for $\varphi \neq 0$ is depicted in
Figs. 4 and 5.
Fig. 4 indicates symmetry breaking for all positive values of $R$ in the
conformal version.
The particular settings in Fig. 5, which depict a given situation
in the general model,
make the passage from symmetric to symmetry-breaking phase
take the specific form of a first-order transition, induced ---of
course--- by the change in curvature. The `broken' phase exists for
every positive $R$ below a critical value $R_c$. Thus, we have found
the existence of a curvature-induced phase-transition in scalar QED
with QG. Examining the RG-improved EP at the critical point, we
obtain the following estimates for the induced Newton
($G_{\mbox{\scriptsize ind}}$) and
cosmological ($\Lambda_{\mbox{\scriptsize ind}}$) couplings:
\begin{equation}
{1 \over 16\pi G_{\mbox{\scriptsize ind}} }
\simeq 4.2 \cdot 10^{-15} \mu^2,
\hspace{1cm}
{ 2 \Lambda_{\mbox{\scriptsize ind}}
\over 16\pi G_{\mbox{\scriptsize ind}} }
\simeq -3.3 \cdot 10^{-6} \mu^4,
\label{estim}
\end{equation}
where, as we have already mentioned, $\mu^2 < \mu_{\mbox{\sz Pl}}^2$.
As one can see, Einstein gravity is induced, with the large cosmological
constant.
In principle, it is not difficult to generalize the above results for
more realistic gauge theories. For example, let us consider the minimal
$SU(5)$ GUT (without fermions) with a 24-plet of gauge bosons
$A_{\mu}$ and a 24-plet of Higgs bosons $\varphi$ transforming
under the adjoint representation of the group $SU(5)$. The Higgs sector
of this theory has the form
\bea
{\cal L}_H&=&\ds -{1 \over 2} \mbox{Tr}
(\partial_{\mu}\phi -ig [ A_{\mu}, \phi ] )^2 \nn
&&\ds -{1 \over 4}f_1 (\mbox{Tr} \phi^2)^2
-{1 \over 2}f_2 \mbox{Tr} \phi^4
+{1 \over 2}\xi R \mbox{Tr} \phi^2 .
\eea
Considering the breaking $SU(5) \to SU(3) \times SU(2) \times U(1)$,
with $\phi= \varphi \left( 1,1,1,-{3 \over 2},-{3 \over 2} \right)$
and working, for simplicity, on flat background, we obtain the
following one-loop effective potential with QG corrections in the
$SU(5)$ GUT (for discussion in curved space with no QG, see \cite{BO}):
\beq
\begin{array}{ll}
V=&\ds {15 \over 16}( 15 f_1 + 7 f_2 ) \varphi^4 \\
&\ds +{15 \over 32 (4\pi)^2}\left[
15 \cdot 64 f_1^2 + 1296 f_1 f_2 + {32 \cdot 91 \over 5} f_2^2 \right.
\\
&\hspace{5em}\ds {375 \over 2}g^4
+\lambda^2\xi^2\left(
15+ {3 \over 4\omega^2} -{9 \xi \over \omega^2}+{27 \xi^2 \over \omega^2}
\right) \\
&\ds\hspace{5em}\left.
-\lambda( 15f_1+7f_2 )\left(
{28 \over 3}+18{\xi^2 \over \omega}-8{\xi\over \omega}-8\xi
+{1 \over 3\omega} \right)
\right]
            \varphi^4\left(
\log{ \varphi^2 \over \mu^2 } -{25 \over 6}
\right) ,
\end{array}
\eeq
where the Landau gauge has been used. This result shows explicitly
the universality of QG corrections and the possibility of
applying them to different theories.

\section{Stability in the Yukawa model with conformally invariant
higher-derivative gravity}

The purpose of this section is to discuss the issue of stability in
a Yukawa theory with conformal higher-derivative QG. The Lagrangian of
the Yukawa theory that we consider reads
\bea
{\cal L}_m&=&\ds
{1 \over 2} g^{\mu \nu} \partial_{\mu}\varphi \partial_{\nu} \varphi
+{1 \over 12} R \varphi^2
-{1 \over 4!} f \varphi^4 \nn
&&\ds +\bar\psi (i \gamma^{\mu}\nabla_{\mu}-h\varphi ) \psi ,
\eea
where $\psi$ is a massless Dirac spinor. It is known that, due to
Yukawa interactions,
the scalar coupling constant becomes negative at large $t$ (high
energies), thus rendering the scalar effective potential unstable.
Here, using the simplest ---\ie conformal--- version of our QG theory
\req{LQG}, we will try to understand whether it is possible to change
the stability properties of the EP to the better side by virtue of
QG corrections.

The RG-improved EP looks like \req{VRGconf}, but
the following subtstitutions must be done for the RG equations
(see \cite{BKSVW,BOS} for QG corrections to Yukawa couplings)
\[
\lambda(t)={ \lambda \over
\ds 1+ { \alpha_1^2\lambda t \over (4 \pi)^2 } }, \hspace{1cm}
\alpha_1^2= {803 \over 60},
\]
\beq
\begin{array}{rcl}
\ds{dh^2 \over dt}&=&\ds{1 \over (4 \pi)^2}\left(
10 h^4 -{61 \over 16} h^2 \lambda \right), \\
\ds{df \over dt}&=&\ds{1 \over (4 \pi)^2}\left(
3f^2 +8 f h^2 - 48 h^4 -{41 \over 8} \lambda f + {5 \over 12} \lambda^2
\right), \\
\ds-{1 \over \varphi}{d\varphi \over dt}&=&\ds{1 \over (4 \pi)^2}\left(
2 h^2 + {27 \over 32} \lambda \right) .
\end{array}
\eeq
In the one-loop approach, the EP is given by
\beq
\begin{array}{ll}
V^{(1)}=&\ds{1 \over 4!} f\varphi^4
+{1\over 48(4 \pi)^2}\left[ {5 \over 12} \lambda^2 - 48 h^4  \right]
\varphi^4
\left( \log {\varphi^2 \over \mu^2 } - {25 \over 6} \right) \\
&\ds -{1 \over 12} R \varphi^2
+{1\over 12(4 \pi)^2}
\left[ 2 h^2 + {27 \over 32} \lambda \right]
R \varphi^2
\left( \log {\varphi^2 \over \mu^2 } - 3 \right) ,
\end{array}
\label{V1Yu}
\eeq
where we have supposed that $f \sim \lambda^2-h^4$ and, therefore,
higher order terms like $f^2$, $f h^2$, $f \lambda$ have been dropped
out. We shall also consider the situation in which $\lambda$ and $h^2$
are of the same order, but $\lambda > 10 h^2$ (otherwise the potential
is necessarily unstable). From expression \req{V1Yu}, it is possible to
get the QG corrections to the scalar-to-fermion mass ratio (the fermion
becomes massive after spontaneous symmetry breaking):
\beq
{m^2(S) \over m^2(F)} =
{ V^{(1) \prime \prime }(\varphi_m)^2 \over h^2 \varphi_m} =
{1\over 6(4 \pi)^2}\left[
{5 \over 12} {\lambda^2 \over h^2} - 48 h^2  \right] .
\eeq
In this model, spontaneous symmetry breaking can take place only
as a result of QG corrections (if $\lambda > 10 h^2$).
On the other hand, Fig.6
illustrates the occurrence of instabilities in this theory, both
with and without gravity.

Analogously, we shall briefly outline how to discuss a variant of the
Standard Model coupled to conformal higher-derivative gravity.
The required one-loop RG functions for such a theory, in the absence of
gravity, were given in \cite{FJSE}(first ref.)--- see also \cite{MV}.
After adding to them the corresponding QG
corrections,
(using the
notations of refs.\cite{FJSE}, but for the scalar
coupling, which now is called $f$),
we are posed with the set of equations:
\[
g^2(t)={ g^2 \over \ds 1 + {19 \over 3 (4 \pi)^2}g^2 t },
\hspace{0.5cm}
g^{\prime 2}(t)={ g^{\prime 2} \over
\ds 1 - {41 \over 3 (4 \pi)^2}g^{\prime 2} t },
\hspace{0.5cm}
g_3^2(t)={ g_3^2 \over \ds 1 + {14 \over (4 \pi)^2}g_3^2 t},
\]
\[
\lambda(t)={ \lambda \over
\ds 1+ { \alpha_2^2\lambda t \over (4 \pi)^2 } }, \hspace{1cm}
\alpha_2^2= {481 \over 30},
\]
\beq
\begin{array}{crl}
\ds{df \over dt}&=&\ds{1 \over (4 \pi)^2} \left(
4f^2 + 12 f h^2 -36 h^4 -9f g^2 -3 f g^{\prime 2}
+{9 \over 4}g^{\prime 4} + {9 \over 2} g^2 g^{\prime 2} +{27 \over 4}g^4
\right.
\\
&&\hspace{3em} \ds \left. -{41 \over 8}\lambda f
+ {5 \over 12} \lambda^2 \right)
\\
\ds{dh \over dt}&=&\ds{1 \over (4 \pi)^2}\left(
{9 \over 2}h^3 - 8g_3^2 h -{9 \over 4} g^2 h
-{17 \over 12}g^{\prime 2}h -{61 \over 16} \lambda h^2
\right) .
\end{array}
\eeq

As shown in Fig. 7,
at $\lambda=0$ ---\ie no gravity---
the scalar coupling can sometimes be negative, but
by virtue of $\lambda$-corrections, the sign can be reversed thus
restoring the stability of the model.
Although the general appearance of this figure is quite similar to one
included in \cite{FJSE}, the effect exhibited is different: there, no
gravity was present and the negativeness could be corrected by rising
the value of $f(0)$;
in our case $f(0)$ is held at a fixed value while we change the strength
of QG perturbations.
In these circumstances, such a stability
restoration can be regarded a purely quantum gravitational effect.

It was pointed out in refs. \cite{DPS}, and first ref. of \cite{FJSE},
that demanding absolute stability of the electroweak coupling implies
that \cite{FJSE}
\beq m_H \ge 1.95 m_t - 189 \mbox{GeV}, \eeq
where $m_H, m_t$ are the Higgs mass and the top mass, respectively.
Thus,
the stability of the scalar coupling constant puts some bounds on the
relation between $m_H$ and $m_t$. As we can infer from the above study,
QG corrections may change or completely destroy these bounds. Of course,
the mechanism of appearance of large QG corrections in the SM beta
functions
is not clear at all, neither do we have reliable estimations for the
QG couplings' initial values. Hence, the above results should be
considered
rather as a speculation which, however, may open quite an interesting
field of QG applications to SM phenomenology.

\section{Discussion}

In this work we have studied the RG-improved (and one-loop) effective
potential on curved background. A general formalism has been applied to
scalar QED with and without $R^2$-gravity, on curved background
for both cases.
We have discussed a few phenomena caused by QG,
particularly dimensional transmutation in the preesence of classical and
quantum gravitational fields, influence of QG on the stability of the
effective potential (in the examples of the Yukawa model and the
Standard Model). In particular, we have shown that, due to QG effects in
the conformal version of $R^2$-gravity, the running scalar coupling may
become less unstable.

A numerical study of the phase structure in scalar QED on curved
background and also in the presence of quantum $R^2$-gravity has been
done. We have shown the possibility of spontaneous symmetry breaking and
curvature-induced phase transitions. Is is interesting that, after
the phase transition, one can get the Einstein theory in the
low-energy limit even in situations where the Einstein sector was not
present in the original $R^2$-quantum gravity.

QG corrections to the scalar-to-vector mass ratio are calculated in
scalar QED. Due to their universal structure, it is not difficult to
repeat the same analysis for any reasonable GUT theory with
higher-derivative gravity.

The general formalism developed in this paper may be easily applied to
any multiplicatively renormalizable theory of matter with QG. There are
many questions pending in such more realistic GUT theories, like
the study of stability in the scalar sector,
development of a clearer understanding of the connections between an
effective theory for QG and GUT phenomenology, the inducing of
Einstein gravity with realistic values of comological and Newton
coupling constants at the critical point of a curvature-induced phase
transition, and so on.

An interesting possibility is connected with an inflationary universe
based on a Coleman-Weinberg type effective potential. One may hope that,
taking into account QG corrections to this potential in the above
discussed form may improve the situation and make such an inflationary
universe more realistic. Viewed from another side, it would be of great
interest to include QG corrections in the back-reaction problem analysis
\cite{FHH} (albeit not so easy). We plan to return to some of these
questions elsewhere

\vskip1cm
\noindent{\large\bf Acknowledgements }

SDO would like to thank I. Antoniadis, M. Einhorn and E. Mottola for
helpful discussions. This investigation has been supported by the
SEP Foundation (Japan), by DGICYT (Spain), by CIRIT
(Generalitat de Catalunya), and by ISF project RI1000 (Russia).

\newpage
\noindent{\large\bf Figure captions}
\vskip1cm

\ni{\bf Fig. 1} RG-improved effective potential for SQED, with
$e^2(0)=0.1$, $f(0)=0.01$, $\xi(0)=0$, ($\mu^2=1$ is used throughout),
for
different values of the curvature $R$. Note the curvature-induced phase
transition, which takes place when $\varphi=0$ becomes
metastable for $R < R_c \simeq 1.44 \cdot 10^{-6}$, while
the local minimum at $\varphi >0$ becomes the global one.

\vskip0.75cm

\ni{\bf Fig. 2} Running scalar coupling $f(t)$, in the
conformal version of SQED with $R^2$-gravity, for different initial
values of $\lambda(0)$. In this and the next figures, the
initial values of $f(0)$ are taken equal to $e^4(0)+\lambda^2(0)$,
for reasons already explained in the text. Since we are
concerned with the gravity-dominated regime, the
value of $\lambda(0)$ has been chosen one order of magnitude larger than
$e^2(0)$.
If, instead of the conformal version, we consider the general theory
with $\xi(0)=1/6$ (and $\omega(0)=1$), the resulting picture
practically overlaps the one here displayed.

\vskip0.75cm

\ni{\bf Fig. 3} $f(t)$ for the general version of the same theory, in
the case $\omega(0)=1$, $\xi(0)=1$.
For $t < t_i \simeq -6$ these couplings become negative.

\vskip0.75cm

\ni{\bf Fig. 4} Effective potentials for the conformal version
of SQED with higher-derivative quantum gravity.
In this range, the one-loop and RG-improved approximations coincide
to the extent that their associated curves completely overlap
one another.
However, closer
numerical examination around $\varphi=0$, at smaller scales than
those visible in the figure, reveals differences between both.
A $\xi(0)=1/6$ choice in the general model would give rise to
practically the same figure as in the conformal case.

\vskip0.75cm

\ni{\bf Fig. 5} One-loop (dashed line) and RG-improved (solid
line) potentials, in the general version of our QG-corrected SQED
model, with
$\omega(0)=1$, $\xi(0)=1$,
$e^2(0)=10^{-2}$, $\lambda(0)=0.15$, $f(0)=e^4(0)+\lambda^2(0)$, for
three different values of the curvature $R$.
At the scale shown, there is already quantitative
disagreement between both approximations in the precise
value of $R_c$, which would be of $\simeq -4.1 \cdot 10^{-9}$ at one loop
and of $\simeq -1.3 \cdot 10^{-9}$ for the RG-improvement, but the nature
of the overall picture still coincides. The existence of a
curvature-induced phase-transition may be inferred from this image.

\vskip0.75cm

\ni{\bf Fig. 6} Running scalar coupling $f(t)$ for the conformal Yukawa
model. The picture represents the cases
$h^2(0)=0.01$, $f(0)=0.01$ and $h^2(0)=0.1$, $f(0)=0.002$. For each
of them, we have considered a pair of curves: (a) in the absence of QG,
(b) with the corrections corresponding to $\lambda(0)=0.25$.
The instability is already noticeable for the second set of initial
values. In the QG-corrected version, $f(t)$ tends to be marginally less
negative than in the absence of gravity.

\vskip0.75cm

\ni{\bf Fig. 7} Behaviour of the running scalar coupling $f(t)$ for
the SM with QG corrections. The present curves correspond to the initial
values $g(0)=0.65$, $g^{\prime}(0)=0.36$,
$g_3(0)=\sqrt{4 \pi \alpha_3}$, $\alpha_3=0.11$,
$f(0)=0.14$, $h(0)=0.7$, and to the different values of $\lambda(0)$
quoted. An increase in this constant may correct the
instability brought about by the negativeness of $f(t)$.


\end{document}